# Einstein and the Solar Eclipses

## Why Albert Einstein joined the Astronomische Gesellschaft in Potsdam 1921

Günther Rüdiger, Leibniz Institute for Astrophysics Potsdam

In 1911 already Einstein suggested that photons which pass close to the Sun are deflected by its effective mass which would apparently shift the positions of stars by up to one arcsecond, something that should be observable by astronomers during solar eclipses. Einstein's ideas including the later doubling of the value due to relativistic space curvature met with enthusiastic responses from astronomers. The history of these efforts, their motivations and interactions with Einstein is presented here, beginning with the expedition that ultimately failed due to WWI of the young Erwin Freundlich in 1914 and ending with the unsuccessful Potsdam campaign to Öland (Sweden) in 1954. A newly found conference photograph from the first post-war meeting of the Astronomische Gesellschaft in 1921 shows Albert Einstein as a new member of the AG on the Telegraphenberg in Potsdam.

## I.

On Monday, 22 August 1921, Albert Einstein wrote to a friend from Haberlandstraße in Berlin, saying that he had spent a happy month with his two boys on the Baltic Sea, first on Fischland and then in Kiel, where they went sailing. The boys, aged 11 and 17, lived with their mother in Zurich for most of the year. Einstein had returned to attend the meeting of the Astronomische Gesellschaft in Potsdam on Wednesday, having recently applied for membership. It was the first meeting since WWI; the astronomers had not seen one another for a long time. The conference took place in the theatre of the Hohenzollern's Stadtschloss on the Alter Markt, where the city parliament had recently begun to meet. Einstein must have been amused by the Prussian Baroque architecture, and also by the very short distance to the city railway station – a palace with a rail link to Berlin; where else would you find such a thing?

Einstein had become an international superstar overnight after – shortly after the war – two British expeditions led by Arthur Eddington had confirmed, during the solar eclipse on 29 May 1919, his prediction that light would be deflected in the vicinity of the Sun. Eddington had seized the opportunity to experimentally answer several fundamental questions in physics at once: How much does light weigh? Is the space curved, as Einstein claimed, or "a result [emerges] of yet more far-reaching consequences – no deflection." The measurements had taken place in West Africa and in Brazil. Telegrams had arrived from Brazil ("Magnificent eclipse") and from the African island of Príncipe ("Cloudy. Full of hope"). Initial reports in September were promising. On 27 September, Einstein had received a telegram from Holland: "Eddington has provisionally found the stellar shift at the edge of the Sun to be between nine-tenths of a second and twice that." Einstein had friends there; the Netherlands had remained neutral during the war, and he had been appointed visiting professor at Leiden University as early as 1920. On the same day, he sent a telegram to his sick mother: "Joyful news today." He immediately published the message in the famous journal Naturwissenschaften: "According to a telegram sent by Prof. *Lorentz* to the undersigned, the British delegation led by *Eddington*, which had been dispatched to observe the solar eclipse on 29 May, found the deflection of light at the edge of the solar disc as predicted by the general theory of relativity. The value determined provisionally so far lies between 0.9 and 1.8 arcseconds. The theory predicts 1.7. Berlin, 9 October 1919. A. Einstein." His information was based solely on hearsay; he was really taking a big gamble and had himself played a major part in stirring up the upcoming spectacle surrounding his person. It must have shaken him up to his core: a few years earlier he had not found a new solution to an old equation, but – only through thinking – had formulated a new equation, and this equation only made sense if the stellar shift at the edge of the Sun really was 1.75´´. Six years earlier, on the occasion of Einstein's appointment as a member of the Berlin Academy, Max Planck had mentioned that the verifiable consequences of the new gravitation theory lay at the limits of what was measurable and hoped that Einstein will develop in Berlin the more important theory of quantum phenomena. But Einstein was determined to test his own theory of gravitation so that he needed the support of astronomers. As early as 1911, in a paper in Annalen der Physik on the influence of gravity on the propagation of light, he had determined the gravitational deflection of light during a close passage by the Sun. He argued that the effect of almost 1 arcsecond calculated with Newtonian gravity should be observable during solar eclipses. Two years later he wrote to Erwin Freundlich in Berlin-Neubabelsberg, saying, “In this matter, astronomers could render an invaluable service to theoretical physics” by observing the eclipse in 1914. He also hoped that the planet Jupiter could also provide measurable shifts of streaking starlight. “As far as Jupiter is concerned, I realise that it is difficult”, he fortified Karl Schwarzschild in Potsdam, but the importance of the new theory requires that “it *must* work.” Many decades later, he was amused by his impatience at the time and stated that the significance of the gravitation theory lay not in the prediction of tiny effects in the Universe but “in the simplicity of its foundations and in its logical consistency”. When asked about God in his world, he wisely replies that God is the plan for the universe which is revealed in the elegance and harmony of the laws of nature.

After years of uncertainty, the decay of his family, serious illnesses and eating disorders, the observational result was that light followed his equation rather than the old one by Newton. As early as 6 November 1919, after a joint meeting of the Royal and Astronomical Societies in London, the Astronomer Royal Sir Frank Dyson announced that the photographic plates of the solar eclipse confirmed Einstein’s prediction that light would be deflected in accordance with his new gravitation law. The New York Times headline immediately read “Lights all askew in the heavens – Einstein theory triumphs”, whilst the London Times declared “Revolution in science, Newtonian ideas overthrown.” In late November, in the London Times, under the rather bold headline “My theory”, Einstein expressed his gratitude to British scientists for having tested an idea that had originated in the midst of the war in the land of their enemies. The compliment is unnecessary, the editor replied coolly. With a one-month delay, the Berliner Illustrirte has published an article entitled “Albert Einstein – A New Figure in World History” followed of hundreds of media tributes, let’s hope that doesn’t backfire, some people must have thought.

The result of the English expeditions had confirmed Einstein’s new law of gravitation, which indeed impressed the Nobel Prize Committee but did not entirely convince them. Eddington had written to Stockholm that Einstein towered above his contemporaries, just as Newton once had, but one committee member considered all the consequences of the new theory to be immeasurably small. The committee awarded its 1921 prize neither to Einstein nor to any other physicist. The confirmation of the revolutionary theory by a single measurement did not yet seem sufficient; they wanted to buy time and await a further experiment. It was true that Eddington’s result was at the very limit of what was feasible; measurements of one-hundredth of a millimetre had to be taken on photographic plates, and there had repeatedly been doubts about the results but the loudest objections came 10 years later from the Einstein Tower in Potsdam.

Einstein had urgently needed the prize money to finance his divorce. As early as 1918, he had notarised a deed awarding the entire sum – which he did not yet even possess – to his wife in Zurich, for her maintenance following the divorce and that of the children. “You would be free to dispose of the interest as you wish. The capital would be deposited in Switzerland and kept safe for the children.” Mileva had returned to Zurich with the boys in 1914 after staying in Berlin for only three months; she could not, or would not, live in the neighbourhood of her husband’s cousin and sweetheart, Elsa Löwenthal. Financial disputes began after only a few days. Your letter, Einstein complained in September, “in which you lament a lack of money, is incomprehensible to me given what I have done.” He had already paid for the move to Zurich, including almost all the furniture. The aged Einstein shall write about a good friend that this man had managed “to live for many years not only in peace, but even in constant harmony with a woman – a challenge in which I have failed twice.”

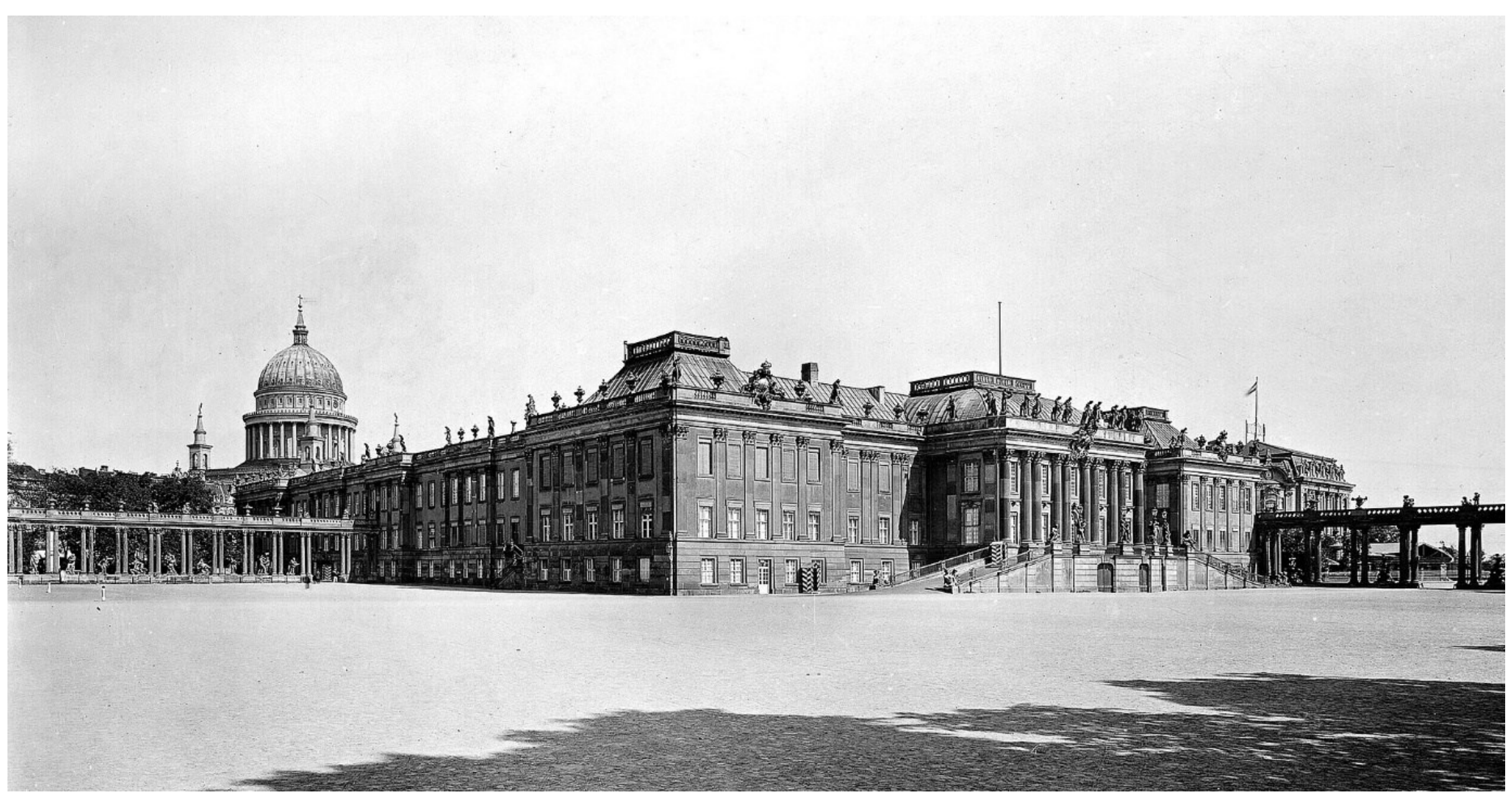

Potsdam Stadtschloss, 1920. It was here that the Astronomische Gesellschaft held its first meeting after the World War, in August 1921. Photo: Wikipedia, public domain

The astronomy conference begins at ten o'clock on Wednesday in the former theatre of the palace. Einstein has appeared in a white summer suit, whilst many participants are dressed in formal black; for the evening, in addition to the conference photograph, a reception is scheduled in the Dome Hall of the Great Refractor on the Telegraphenberg. "Allow me to compare your conference to one of the first budding twigs on the tree of science," says the State Secretary of the Prussian Ministry of Science, opening the first session of the astronomers since the great World War. For Elis Strömgren, the Danish chairman of the Astronomische Gesellschaft, "Potsdam holds a special appeal thanks to its large, world-famous Astrophysical Observatory." A record-breaking 150 participants from 14 nations have gathered, including Arthur Eddington as a member of the AG, who, against all national opposition, has come to Potsdam to 'heal the wounds of war'. The victor alliance had banished German scientists from international cooperation and boycotted all specialist conferences, as well as German as a language of science. Eddington didn't care about that; as a pacifist and Quaker, he had other ideas: he wanted to meet Einstein in Potsdam and to have his own results verified. Even during the war, Einstein had suffered from the isolation of scientists, and he had hoped for an initiative from those who had gained 'a superior reputation through intellectual achievements' in the civilised world. The political firewall had held until Germany joined the League of Nations in 1926, but had not hindered developments such as those in quantum theory at German universities but not in Berlin.

On the very first day, the commission that was of highest importance to Einstein was established. In a proposal to the executive committee of the Astronomische Gesellschaft, Einstein, Eddington and other astronomers had written that the next total solar eclipse would take place in the Indian Ocean in September 1922, which could be used to verify the results from 1919. Indeed, an organising committee was appointed, comprising Einstein, Freundlich and Director Ludendorff; the Astrophysical Observatory had already applied for the necessary funding. As German inflationary currency was virtually worthless abroad, the project was to be a joint German-Dutch endeavour. The committee would later appoint Erwin Freundlich from Potsdam as head of the German expedition. This was an obvious choice, given that the young Freundlich – not yet thirty years old – had been the first astronomer in the world to attempt, during a solar eclipse in the Crimea in 1914, to measure the deflection of starlight as it passed the Sun, a phenomenon proposed by Einstein three years earlier. Freundlich had studied mathematics in Göttingen, but had only found a minor assistant post in an insignificant field of research at the Sternwarte Berlin-Neubabelsberg. He was travelling through Europe with his bulky 3.4-metre astrograph in tow – a highly complex organisational problem

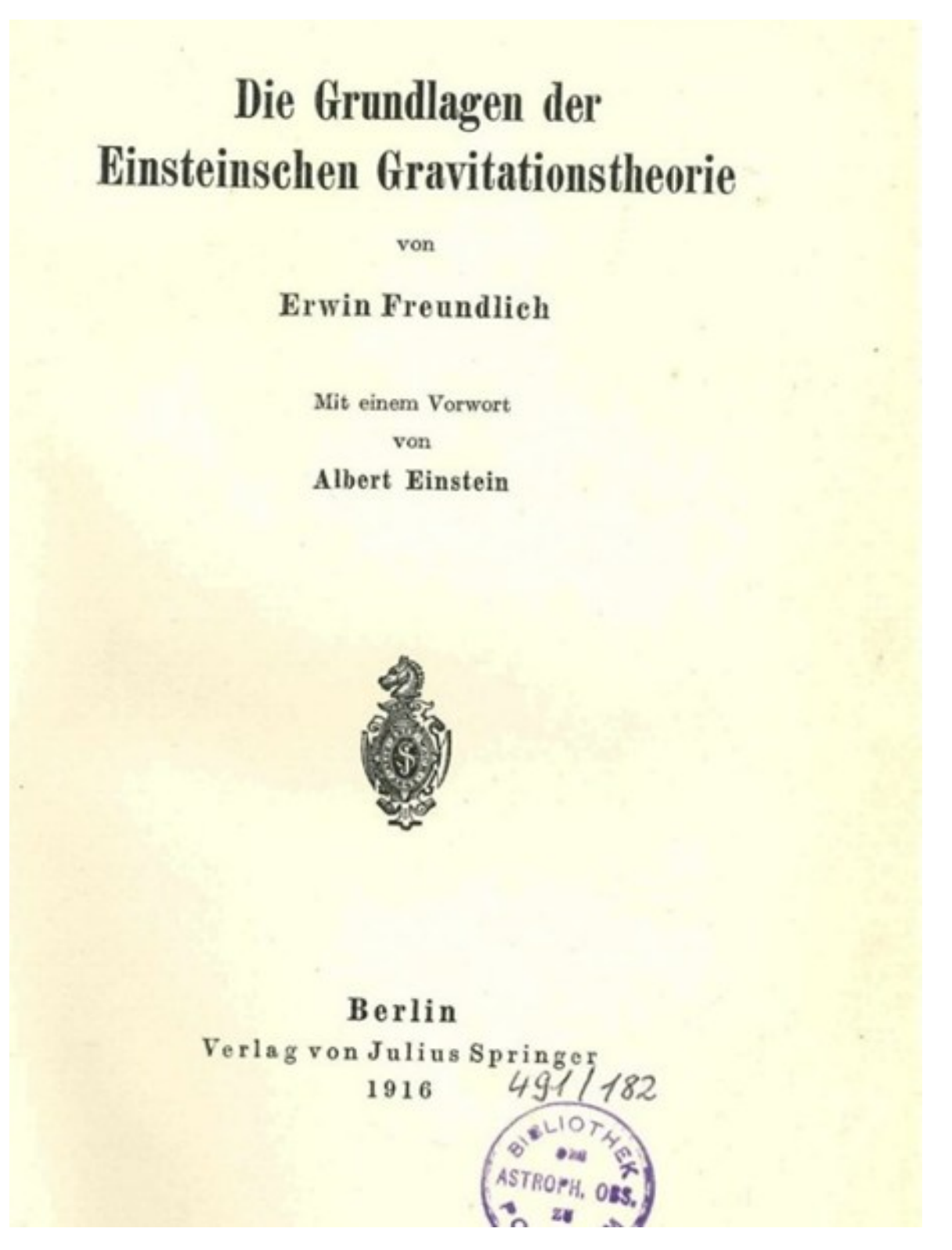
Die Grundlagen der
Einsteinschen Gravitationstheorie

von

Erwin Freundlich

Mit einem Vorwort
von
Albert Einstein

Berlin
Verlag von Julius Springer
1916

The mathematician Erwin (Finley) Freundlich had obtained his doctorate in function theory in Göttingen in 1910 and was a very early ‘Einsteinian’.

financed by external sponsors. Shortly before his departure, he had quickly calculated whether Einstein’s idea might not also be tested with a good stopwatch during the occultation of a star by the Moon near the Sun. Einstein, who had set Freundlich on this path, wrote to a friend: “I no longer doubt the correctness of the whole system, whether the observation of the solar eclipse succeeds or not.” The outbreak of WWI destroyed all the efforts and Freundlich was interned in Odessa, later being exchanged as part of a military prisoner swap. The astrograph was lost. At the same time, Adolf Miethe, professor of photography and chemistry of the Technische Hochschule Charlottenburg just a few miles from Potsdam, had started a well-equipped expedition for northern Norway, where he experienced his first solar eclipse in perfect weather. It is challenging to imagine what might have happened had Freundlich joined Miethe: the young assistant at the Sternwarte Berlin-Babelsberg would have been the first to find the mass of light. Still on the basis of the Newtonian gravitation theory Einstein had predicted only half the actual deflection of light. How might he have commented on that? Everything would have turned out differently – with the present text, too.

Late on Wednesday afternoon, the conference participants visited the Astrophysical Observatory on Telegraphenberg. Freundlich gives a short demonstration of his new tower telescope, which didn't play an important role at the conference, but which even later could hardly contribute to its intended purpose of 'the experimental development of the theory of relativity'. An exhibition of the latest models of mechanical and electrical calculating machines in the corridors of the main building made the greatest impression. Many visitors would never have seen these wonder machines before; it was their first encounter with the coming revolution in computing technology.

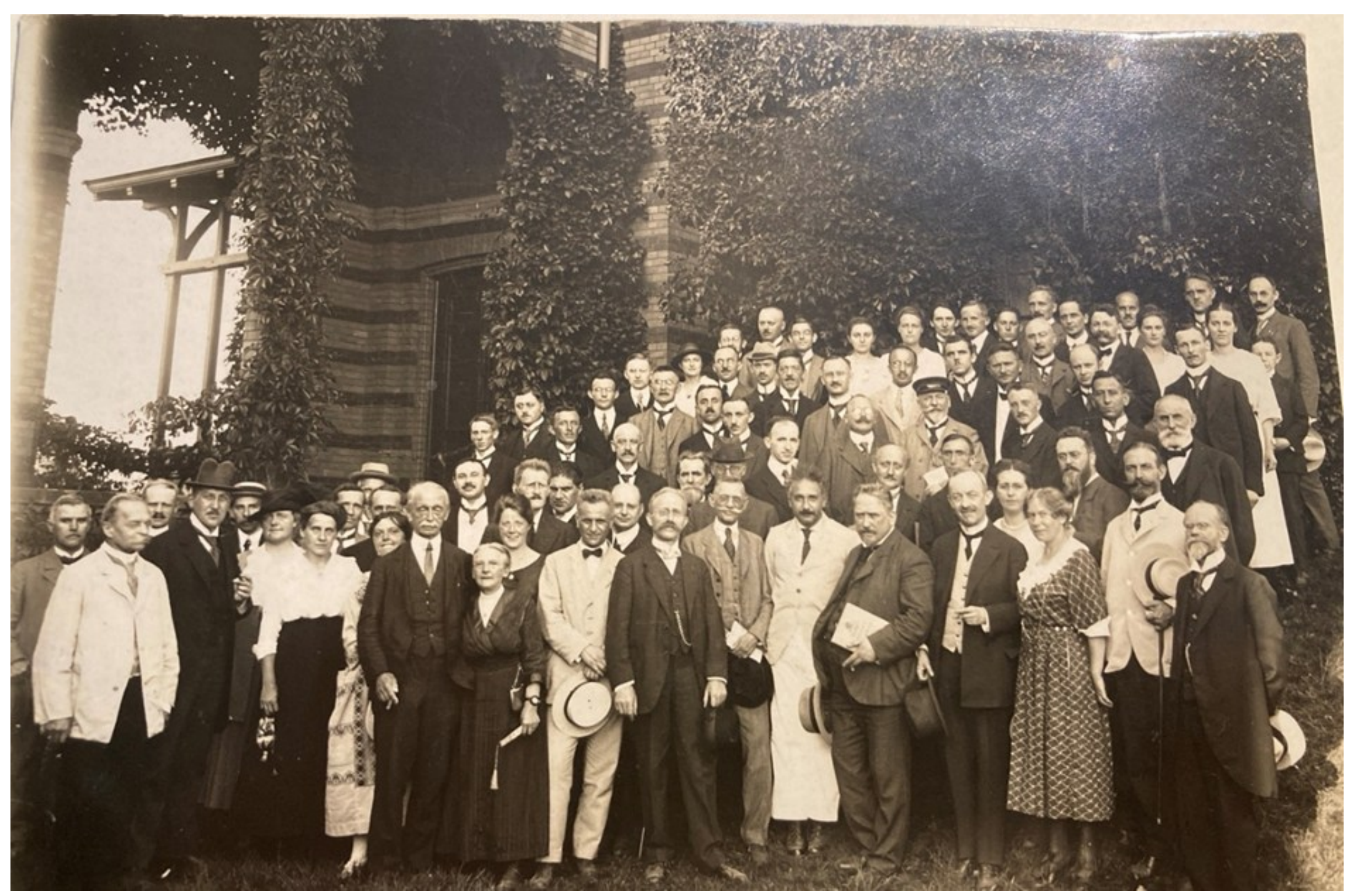

Conference photo 24 August 1921, Telegraphenberg. In foreground: Einstein, Strömgren, Ludendorff (from left); Eddington and Freundlich behind Einstein. From the estate of Walter Hassenstein, courtesy of Wolfgang Hassenstein

II.

The 1922 solar eclipse expedition turned out to be a real nightmare. The two enormous instruments from Potsdam, with tube lengths of up to 8 metres, had to be shipped from Hamburg to Jakarta/Indonesia taking six weeks. This was followed by a three-day stormy crossing in a light boat to the unfriendly Christmas Island; each photographic plate weighed 3 kg on its own! The weather on the island, however, had been much-promising right up until 21 September, with consistently clear skies; but just as the Moon first touched the Sun, the sky had clouded over. The team were only able to take a few images of the solar corona through gaps in the clouds, but not a single exposure to record the positions of the stars near the Sun.

The Nobel Prize Committee eventually took pity on Einstein after all. On 9 November 1922, the world's most famous physicist was retrospectively awarded the prize for 1921 – cautiously not for the theory of relativity but for another of his creations. On 10 December, just as the unsuccessful German astronomers had returned home from Christmas Island, the Nobel lecture was delivered in Einstein's absence. However, in June 1923, the leaders of the solar eclipse expedition from the Lick Observatory announced the results of their measurements taken on 21 September 1922 on the west coast of Australia under ideal weather conditions: one observer had obtained an angle of 1.6 arcseconds, another 1.8. Fortunately, the analysed stars showed an optimal distribution of the distances from the Sun so that internal errors remained small. "Our observations agree very closely with Einstein's prediction." Only now did the German antisemitic and nationalistic hostility – even led by one or two Nobel laureates in physics – die away. If my theory had turned out to be wrong, Einstein sarcastically wrote, France would have said that I was German and Germany will declare that I am a Jew.

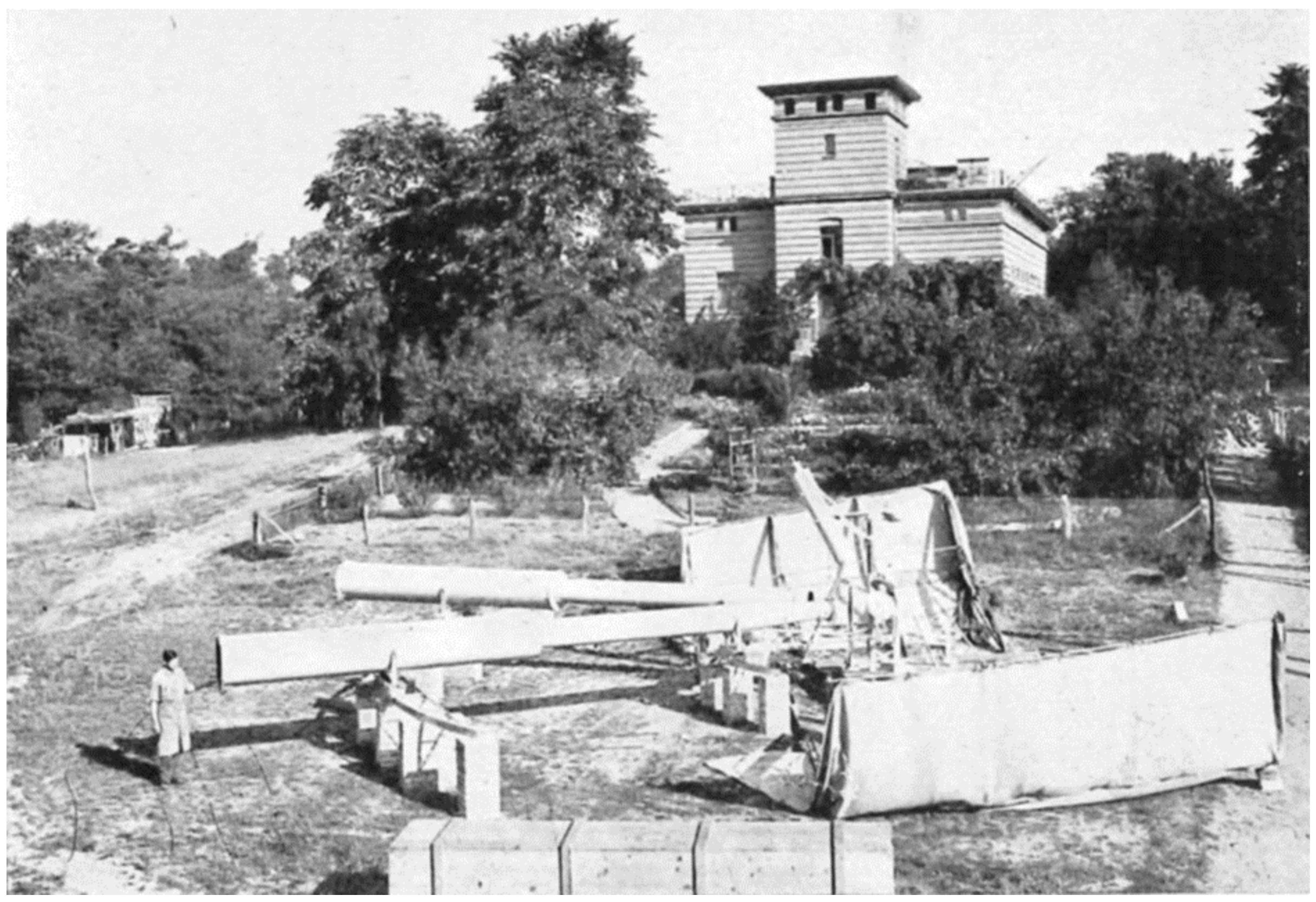

The Potsdam double horizontal camera from 1929, with a heliostat and deflecting mirrors, focal length 8.6 m.

It was much later that Erwin Freundlich was able to collect his own data during an eclipse expedition to Sumatra organised by the Astrophysical Observatory Potsdam; analysing this data would occupy him for the rest of his life, even though the subject had, in fact, been considered closed. His attempts in 1922 and 1926 had proved fruitless, but on 9 May 1929 the happy moment had finally arrived: for the first time in 15 years, he was in the right place. “Fortunately, this time the sky cleared completely even before totality set in. Shortly before the Sun was completely eclipsed by the Moon, the lighting conditions were so unusual that one lost all sense of the state of the sky. Many bright stars became visible.” The exposures, again taken with a 3.4-m astrograph and the now perfected giant Potsdam double horizontal camera, were successful. The four exposed plates of the Sun’s surrounding and that of a control field were reduced in Potsdam in the following years. The constellation of the background stars in the Sun’s vicinity, however, was less favourable than it had been in 1922. The value of the light deflection now appeared to be highly precise, but was 30% above the relativistic value; it was announced that the observations were 'incompatible with the value claimed by the theory'. Ludendorff published a personal reduction of the Sumatra data leading to 1.90´´ – much closer to Einstein's value.

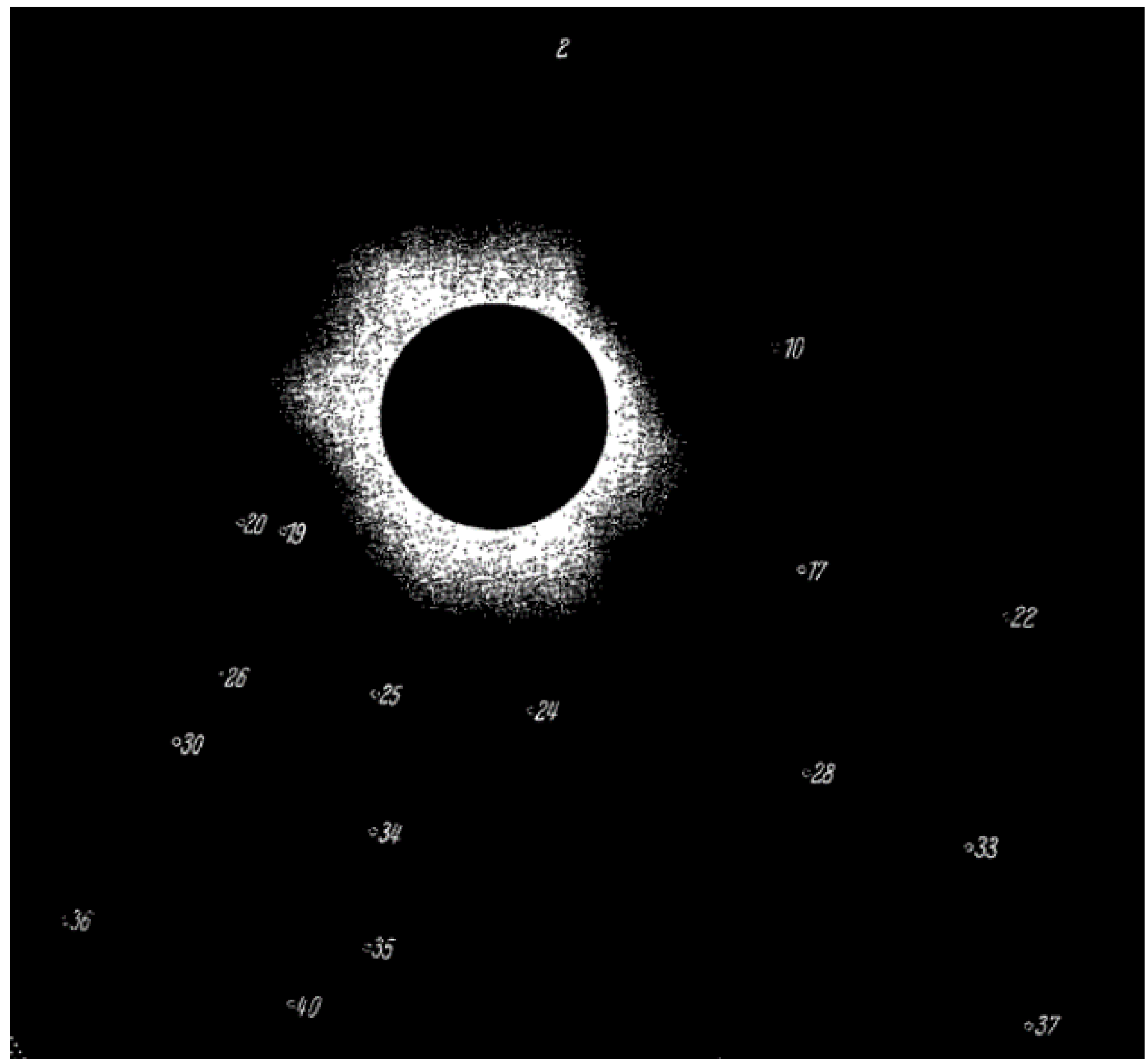

One of the successful images taken with the Potsdam double horizontal camera on 9 May during the expedition to Sumatra. The key background stars are marked in the original paper.

At the Einstein Tower, however, there was firm belief in the own results hence Eddington's 1919 analyses were loudly and publicly called into question, even in Einstein's presence. He, who as chairman of the Einstein Tower's advisory board formally was Freundlich's superior, had informed Max von Laue in those days, saying that he was always looking forward to the meetings on the Telegraphenberg because they were always lively affairs soothing his 'dark soul'. It was now by no means clear whether Freundlich intended to confirm or reject Einstein's theory; the temptation

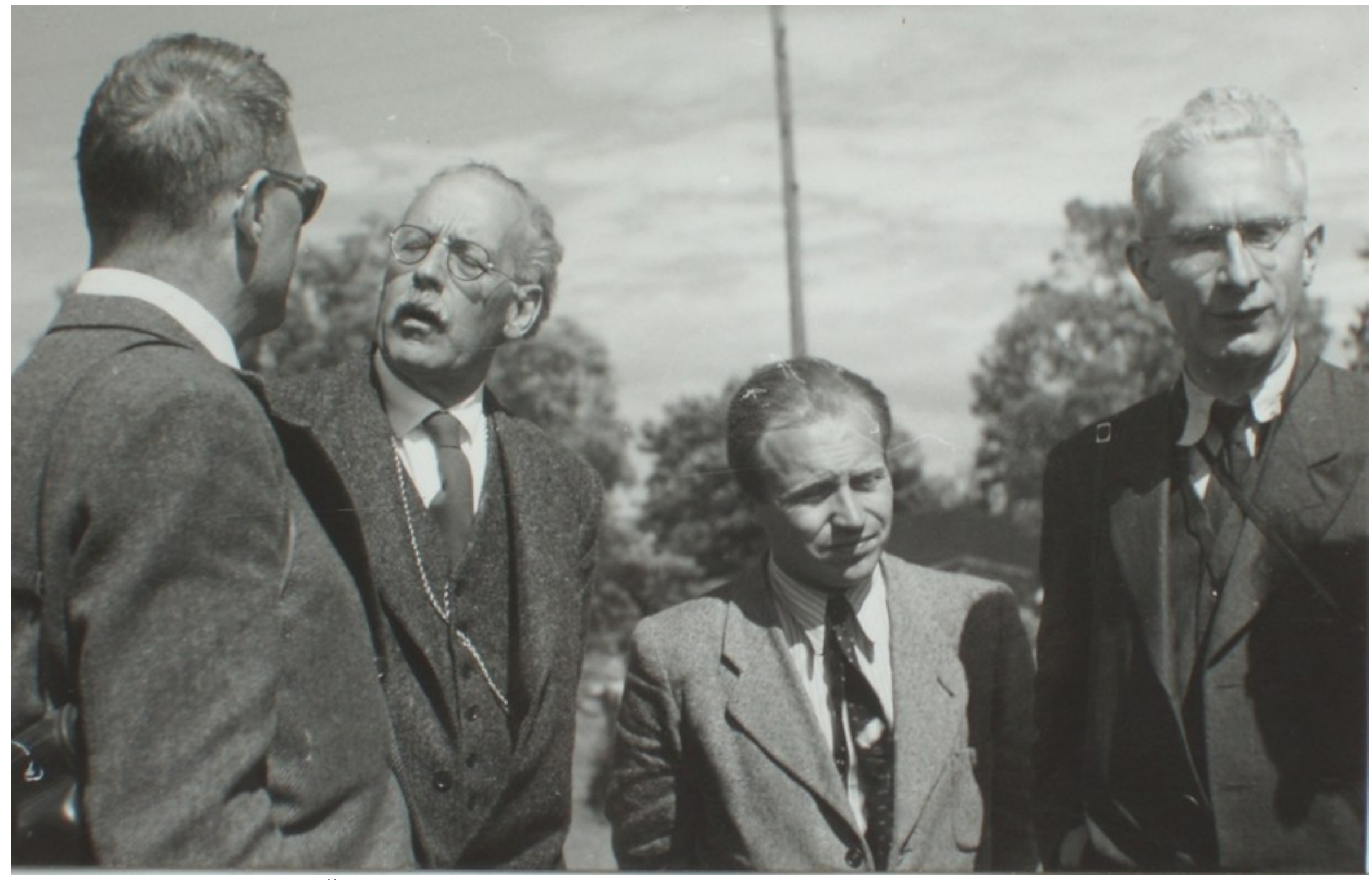

Eclipse expedition to Öland (Sweden) in 1954. From left: Schürer, Freundlich, Mattig, Wempe. Mattig estate, courtesy of W. Schmidt.

to make history as an astronomer was immense. In August 1951, Freundlich travelled from Scotland to East Berlin to secure the GDR government's support for the Potsdam Observatory to verify his old results on light deflection during a solar eclipse expedition to Sweden or Norway in 1954. He argued that it would be of the highest importance for the experiment to be repeated using the old horizontal double camera which should still exist. Moreover, at the 1952 Physics Congress in Berlin, he insisted on the findings of the Sumatra expedition and postulated an as yet unknown alteration of light as the cause when the object passed close to the Sun. Because of the upcoming solar eclipse, the Berlin Academy of Sciences consulted Einstein in Princeton, and he replied that he doubted the profit of such an expensive project. Previous expeditions have already achieved the maximum possible accuracy for such observations. Even Grotrian as the director of the Observatory didn't want to update the giant camera system, but he had to do it and the camera was fully overhauled in Jena. The expedition to Öland took place with Freundlich participating, but remained without any result due to bad weather on 30 June. Later, radio-astronomical methods were used to confirm, with a high degree of precision, the value calculated by Einstein for the deflection of light by the Sun, even in the absence of any eclipse.

## III.

Einstein was never again seen at an astronomical conference; the meeting in Potsdam remained the only one he ever attended. On the eve of his final departure from Germany, on 6 December 1932, he ended his relation with Potsdam with a letter about a violin from that city. ‘I had the opportunity today to try out a violin made by Erich Kilow. Mr Kilow’s violin is undoubtedly one of the finest violins I have ever held. It responds easily and has a round, balanced tone. There is no question that such artistic skill in violin-making deserves to be promoted; we must finally move beyond the stage where people think that an excellent violin must be old.” It’s always different from what you think, is what it probably was meant to mean; he would never set foot on German soil again. Less than a year later, his former favourite and most recent rival, Freundlich, was also forced to escape the Nazi power in Potsdam after director Guthnick of the Sternwarte Berlin-Babelsberg had branded him in a letter of July 1933 (!) to the Ministry of Science as a restless “fighter of boosting the Jewish influence in German cultural life”.